\documentstyle[aps,prl,epsfig,twocolumn,times]{revtex}

\renewcommand{\vec}[1]{{\bf #1}}

\newcommand{\be}{\begin{equation}}
\newcommand{\ee}{\end{equation}}
\newcommand{\bea}{\begin{eqnarray}}
\newcommand{\eea}{\end{eqnarray}}

\newcommand{\PRB}[3]{Phys. Rev. B {\bf #1}, #2 (#3)}
\newcommand{\PRL}[3]{Phys. Rev. Lett. {\bf #1}, #2 (#3)}
\newcommand{\ea}{\mbox{\em et al.}}

\begin{document}
\draft
\def\overlay#1#2{\setbox0=\hbox{#1}\setbox1=\hbox to \wd0{\hss #2\hss}#1%
\hskip-2\wd0\copy1}

\twocolumn[\hsize\textwidth\columnwidth\hsize\csname@twocolumnfalse\endcsname

\title{ 
Effects of Electronic Correlations on
the Thermoelectric Power of the
Cuprates}
\author{G. Hildebrand, T.J. Hagenaars, and W.~Hanke}
\address{Institut f\"ur Theoretische Physik, 
Universit\"at W\"urzburg, Am
Hubland, 97074 W\"urzburg, Germany}
\author{S. Grabowski, J. Schmalian\cite{joerg}}
\address{Institut f\"ur Theoretische Physik,
 Freie Universit\"at 
Berlin, Arnimallee 14, 14195
Berlin, Germany}

\maketitle
\begin{abstract}
We show that important anomalous features 
of the normal-state   thermoelectric power $S$
of high-$T_c$ materials can be understood as being caused by doping 
dependent short-range antiferromagnetic correlations. 
The  theory is based on  the fluctuation-exchange 
approximation applied to Hubbard model in the 
framework of the Kubo formalism.
Firstly, the characteristic  maximum of $S$ 
as function of temperature can  be explained by the
anomalous momentum dependence of the single-particle 
scattering rate.  
Secondly, we  discuss the role of the actual
Fermi surface shape for the occurrence of a sign
change of $S$ as a function of temperature
and doping. 
\end{abstract}
\pacs{PACS numbers: 71.27.+a, 74.25.Fy, 74.20.-z, 74.72.-h}
\vskip0pc]

The thermoelectric power (TEP), $S$,  of the high-$T_c$ materials 
in the normal state
exhibits anomalous 
features\cite{OCT92,BernhardTallon,TallonCooper95,Zhou95} 
that are not well understood at present.
For example, its temperature dependence  shows a characteristic 
maximum that can 
be found in all optimally and underdoped  cuprates, which 
is in contrast to the conventional linear-$T$ behavior 
of a weakly correlated Fermi liquid.
Furthermore,  a better understanding of 
the doping dependence,  which has been  shown to be universal 
for a large class of high-$T_c$ materials\cite{OCT92}, is needed.

Experimentally, the TEP
is positive for the {\em underdoped} cuprates. 
It increases rapidly as a function of
temperature,
reaches a maximum $S^\star$ at a temperature $T^\star$ and 
falls off almost linearly with temperature. The size of the
maximum $S^\star$ and the value of $T^\star$ decrease rapidly
with increasing doping, while the slope for the fall-off at
$T>T^\star$ is almost independent of the doping
concentration. The TEP of nearly {\em optimally doped} materials 
shows a similar behavior  below room temperature,
but the over-all size of the TEP is reduced. Furthermore,
the TEP for most of the 
optimally doped samples changes sign at approximately room
 temperature\cite{OCT92}.
For {\em overdoped} samples,
the behavior of the TEP is very different: It  is  negative and 
decreases linearly with increasing temperature.   
At present there are only two classes of high-$T_c$ materials that
do not follow this generic
trend, namely   $\rm La_{2-x}Sr_xCuO_4$ (LSCO) and
$\rm YBa_2Cu_3O_{7-\delta}$ (YBCO). For LSCO the TEP 
remains positive
in the overdoped regime, where it exhibits a maximum  of decreasing 
height for increasing hole concentration \cite{Zhou95}, whereas
YBCO shows a negative TEP in the overdoped regime but with a
positive slope at temperatures 
$T > T^\star$~\cite{OCT92,BernhardTallon}.
Recently Bernhard and Tallon~\cite{BernhardTallon} 
presented strong experimental evidence that the chain contribution 
is responsible for this non-generic TEP  of YBCO while
the contribution
of the $\rm CuO_2$ planes
follows the generic behavior.

Theoretically,  important open problems concerning the 
 characteristic behavior 
of the TEP remain,  although there have been
several attempts to explain this behavior, ranging from 
the van Hove scenario \cite{NewnsT94,McIntosh} to 
phonon drag effects \cite{Trodahl}.
The  most important open questions are the physical origin of the 
characteristic temperature scale $T^\star$ and of the doping 
dependence of the sign of $S$.

In this paper, we demonstrate that short-range
 antiferromagnetic correlations 
are responsible  for salient features of the temperature and
 doping dependence of the TEP.
In particular, we show that for underdoped 
systems the occurrence of a pseudo-gap
in the single-particle excitation spectrum 
 and for optimally doped systems
the pronounced increase of the scattering
 rate   near the "hot-spot"
regions (near $(\pi,0)$) are essential. 
As was proposed by Stojkovi\'c and 
Pines~\cite{SP96}, the latter
phenomenon is also of importance for the
 description of the temperature
dependence of the  Hall coefficient.
Furthermore, we discuss the influence of the shape 
of the Fermi surface of LSCO and YBCO and 
of the bilayer  
coupling.

We note that antiferromagnetic spin correlations
have recently also been argued to be intimately
related to the normal-state pseudogap
in underdoped cuprates and to a sizeable deformation of the
quasi-particle band structure 
and Fermi surface as a function of doping \cite{PGAF,BerlinPRL}.
These ideas have been corroborated with evidence from
Quantum-Monte-Carlo (QMC) simulations in a recent work
by Preuss {\em et al.}\cite{PreussPRL}.
These latter results also provide numerically, in principle  exact
support for the relevance of antiferromagnetic
correlations that are found to also be crucial for the
thermoelectric power in the theory presented here.

The thermoelectric power $S$ is determined by 
the relation between the 
electrical current $\vec{j}$ and the temperature 
gradient $\vec{\nabla} T$:
$
\vec{j}= \sigma_{\rm dc} \left( \; \vec{E} -  S \;   
\vec{\nabla} T   \right)
$,
where $\sigma_{\rm dc}$ is the dc-conductivity
 and $\vec{E}$ an applied electrical field.
In order to account for the strong electronic 
correlations and 
the pronounced short-range  antiferromagnetic 
correlations of the cuprates, we
use the two-dimensional single-band  Hubbard model  
with $ \epsilon(\vec{k}) = -2 t  \left( \cos k_x  + \cos k_y \right) 
-4 t'  \cos k_x   \cos k_y  
-2 t'' \left( \cos(2 k_x) + \cos(2 k_y) \right)\;$
 for  the  microscopic calculation of $S$.
We set $t= 0.25 {\rm eV}$ and $U=4t$.
The values of the longer-range transfer integrals will be
given  below. 
The  transport coefficients  $\sigma_{\rm dc}$ and $S$ are
calculated within the Kubo-formalism as
\bea
 \sigma_{\rm dc} &=& e^2 \int {\rm d}\omega \left(-\partial f(\omega)/\partial
\omega \right) \sigma(\omega )\label{eq:L11} \\
 S&=& -\frac{e}{\sigma_{\rm dc} T} \int {\rm d}\omega  \left(-\partial
 f(\omega)/\partial
\omega \right)  \omega \; \sigma(\omega) 
\label{eq:L12}
\eea
with the differential conductivity 
\be
\sigma(\omega )= \frac{2\pi}{N} \sum_{\vec{k}} \; v^2_{\vec{k},x} \;
 A^2(\vec{k}, \omega ) \, .
\label{eq:sigma}
\ee
Eqs.~(\ref{eq:L11}) and (\ref{eq:L12}) are obtained following,
for example, the
  derivation of Moreno and Coleman \cite{MorenoColeman} and 
approximating the full vertex function by the velocity
$v_{\vec{k},x} = \nabla_x \epsilon(\vec{k})$.
Here $e$ is the electronic charge, $N$ the number of sites,
$f(\omega)$  the Fermi function 
and $A(\vec{k}, \omega )=-\frac{1}{\pi}
{\rm Im } G(\vec{k}, \omega )$  the spectral function.
The single-particle propagator $G(\vec{k},\omega)$ 
 is  calculated within 
 the fluctuation-exchange approximation
(FLEX)\cite{FLEX}. This
approximation
includes the interaction of the electrons with charge and spin
fluctuations.
We employ a 
real-frequency approach to this approximation,
which directly yields $A(\vec{k}, \omega )$ \cite{reelleFLEX}. 
%
%
\begin{figure}[htb]
\centerline{\psfig{file=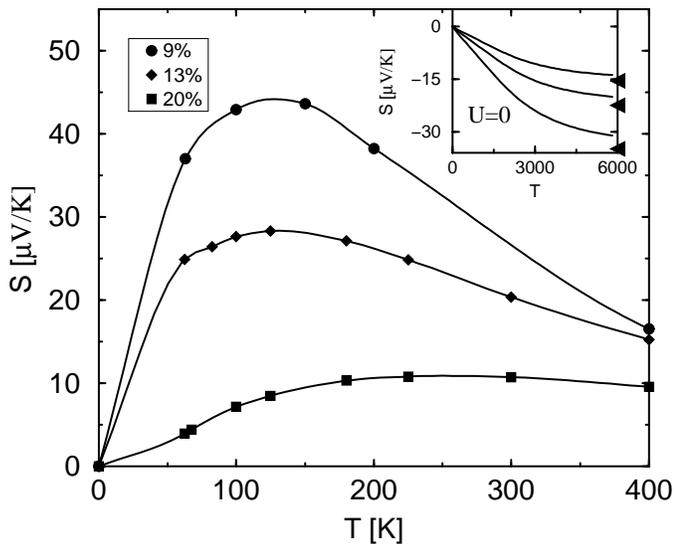, width=8.5cm,%
bbllx=52,bblly=24,bburx=540,bbury=442}}
\caption{\label{fig:lsco_temp} Thermopower $S$ for the LSCO system as a
function of temperature for different doping concentrations. The inset
shows $S$  of  the tight binding model for the same  
doping values. 
The  triangles  indicate the high-temperature
limit given by the generalized Heike formula.}
\end{figure}

In Fig.~\ref{fig:lsco_temp}, 
we show  the TEP for the single-layer LSCO system
 as  function of temperature for 
different doping levels. Here  LSCO is described by 
$t'=t''=0$ to ensure that the   Fermi
surface   is closed around $(0,0)$~\cite{Howell}.
For all fillings, we  find the interesting
result that our theory yields a
positive TEP 
which decreases with increasing hole concentration
and shows pronounced maxima at about 150K for 
the two lowest doping levels.
For the higher doping level (20\%), the maximum is very broad and shifts to
higher temperatures. This  is very similar to the 
experimental observation  in LSCO\cite{Zhou95,CooperLoram96}.
Note that also the absolute values of the
calculated TEP are  in good agreement with experimental 
data~\cite{Zhou95}.  
 
In order to understand the  role of electronic 
correlations for the $T$-dependence and sign 
of  $S$ and to relate them to previous theories,
it is at first  of interest to contrast them   with the TEP of the 
noninteracting system ($U=0$). 
Here $S$ is obtained    from a   tight-binding model
calculated in the relaxation-time approximation (the $U=0$ calculations
along the lines of eqs.~(\ref{eq:L11}) to (\ref{eq:sigma}) yields a
vanishing TEP) with a momentum and
frequency independent relaxation time, see the inset of
Fig.~\ref{fig:lsco_temp}.
In this approach, $S$  is
negative
and proportional to $T$
in the relevant temperature range
 ($T < 600$ K). 
In the high temperature limit, the TEP  
deviates from this linear behavior for $T\sim t$ and
finally  approaches the values given by a 
generalized high-temperature Heike's 
formula\cite{ChaikinBeni}.
Such a $U=0$ theory, in which no details of the relaxation
or scattering rate
are included, still fails to describe
experiments even when a more sophisticated 
phenomenological band structure with a pronounced flattening
of the dispersion near the Fermi level \cite{Putz} as found in 
photoemission experiments~\cite{ARPES,footnote} is included.
Thus, the strong qualitative 
disagreement between the above $U=0$ results and our FLEX results 
for the interacting system implies that for an understanding
of the experimental data it is necessary to
take into account the 
momentum and frequency dependence of the scattering rate.
In order to demonstrate  in more detail 
the role of electronic correlation effects
for $S(T)$ in our approach, we now show $(i)$ that our results  
for the sign of $S$  are caused
by the anomalous  $\vec{k}$ dependence of the scattering rate 
$1/\tau(\vec{k})=-\mbox{Im }\Sigma(\vec{k}, \omega =0)$ 
(where $\omega =0$ refers to the Fermi energy) due 
to antiferromagnetic correlations and  
$(ii)$ that   the temperature dependence of $S$ is determined
by the characteristic  temperature scales of these correlations.
%
%
%
\begin{figure}[htb]
\centerline{\psfig{file=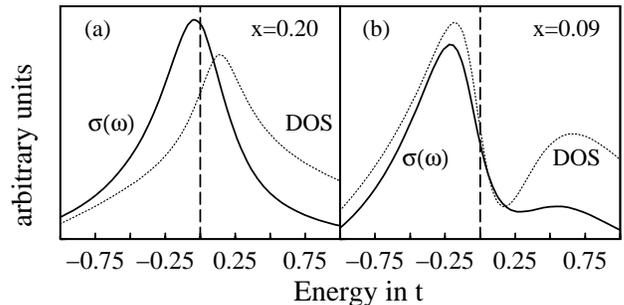, width=8.5cm}}%
\caption{\label{fig:diff_cond}
Differential
conductivity (solid lines) and density of states (dotted lines) 
for two doping  levels  and
 $T=200\,{\rm K}$. All for $U=4t$.}
\end{figure}
It is instructive to consider 
the differential conductivity $\sigma(\omega)$
of  Eq.~(\ref{eq:sigma}), since the TEP, from Eq.~(\ref{eq:L12}), is a measure
of the asymmetry of
$\sigma(\omega)$ with respect to  $\omega=0$ 
for energies
of the order of  $  k_{\rm B} T$.
Thus the TEP is very sensitive to  excitations of states which are
close to but not directly at the Fermi surface.

\begin{center}
\begin{figure}[htb]
\psfig{file=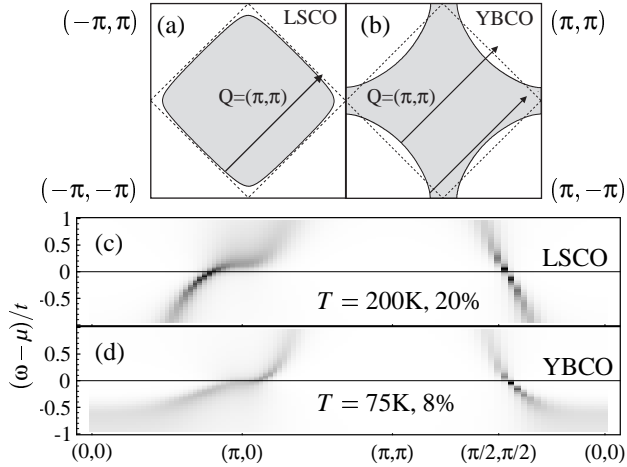,width=8.5cm}
\caption{\label{fig:ako} Fermi surface, (a,b), and spectral function, 
(c,d), for the LSCO and YBCO  parameters, respectively. 
In (c) and (d)
dark (white) areas correspond to large (small) spectral
weight.}
\end{figure}
\end{center}

In Fig.~\ref{fig:diff_cond}(a), 
we show $\sigma(\omega)$ and the density of states
$\varrho(\omega)=\sum_{\vec{k}} A(\vec{k},\omega)/N$ 
for a
doping concentration of 20\%:
Although $\varrho(\omega)$ is peaked for $\omega > 0$ (as in the
uncorrelated case), the differential conductivity $\sigma(\omega)$
exhibits its maximum for $\omega < 0$, yielding the positive value
of $S$.
This behavior can be understood if one takes into account that in 
Eq.~(\ref{eq:sigma}) momentum states with small
velocity $\vec{v}_{\vec{k}}$ and large line width 
(large scattering rates) are suppressed.
Antiferromagnetic spin fluctuations lead
 to large scattering rates for momenta near 
 ${\vec k}_{\rm F} + {\vec Q}$
(${\vec k}_{\rm F}$ Fermi momentum, ${\vec Q}=(\pi,\pi)$) 
which, for LSCO, are everywhere outside of the Fermi
surface (see Fig. ~\ref{fig:ako}(a)).  
%
%

Thus
the  scattering rate of
unoccupied states (holes) near the Fermi surface is 
 larger than the scattering rate of 
the corresponding occupied states (electrons).
This suppresses the contribution of the holes (at $\omega > 0$) to
$\sigma(\omega)$, which  as a result has a
maximum   at $\omega < 0$.
This effect is illustrated in detail in Fig.~3(c), where we show the 
spectral function $A(\vec{k},\omega)$ in the vicinity of $\omega=0$.
It can be clearly seen that the spectral weight
is more sharply peaked for occupied than for unoccupied states around
$(\pi,0)$.
The  suppression of states near  $(\pi,0)$ due to the 
low velocities near the van Hove singularity  also contributes to 
the absolute magnitude of $S$ but does not determine its sign
and order of magnitude.
This demonstrates that the thermoelectric power is determined by  
the combined momentum and energy 
dependence of the scattering rate, making transport for
holes  less coherent than for electrons, yielding $S>0$. 
The transition to conventional  behavior  
occurs at temperatures for which  
the antiferromagnetic correlation length becomes smaller than
the range of the hopping matrix (determined by $t,t'$ and $t''$),
such that no 
pronounced momentum dependencies in the scattering rate
occur~\cite{SP96,PreussPRL}.
For decreasing doping concentration, the
 strength of the antiferromagnetic
correlations increases, leading
to the appearance of a pseudogap in the DOS as a precursor
of the antiferromagnetic phase transition~\cite{BerlinPRL,PreussPRL}.
The FLEX calculation for $U=4t$ \cite{BerlinPRL}
yields results for the
pseudogap qualitatively similar to those obtained in
numerically exact QMC simulations~\cite{PreussPRL}.
The density of states showing the pseudogap is presented 
in Fig.~\ref{fig:diff_cond}(b).
The asymmetry in $\varrho(\omega)$ caused by the antiferromagnetic
fluctuations is even more 
pronounced in  $\sigma(\omega)$ such that 
these correlations
cause in addition to the positive sign a pronounced increase of 
the magnitude of $S$.
Even more interesting, the $T$-dependence of the  
antiferromagnetic fluctuations, which also manifests itself in the 
 pseudogap \cite{BerlinPRL}, determines the anomalous  
$T$-dependence of the TEP in the under- and optimally doped systems of 
Fig.~\ref{fig:lsco_temp}.
 Note, however, the weak  doping-dependence of $T^\star$ in our theory.
We believe
that this shortcoming  is related to the difficulties in the theoretical description
of the single-particle excitation spectrum of underdoped systems rather
than to the approximate
treatment of the transport problem.
We believe that  a  better description of the detailed momentum dependence of the pseudogap,
in agreement with recent angular resolved photoemission 
experiments~\cite{ARPES} and QMC results~\cite{PreussPRL} is essential here.
%
%
\begin{figure}[htb]
\centerline{\psfig{file=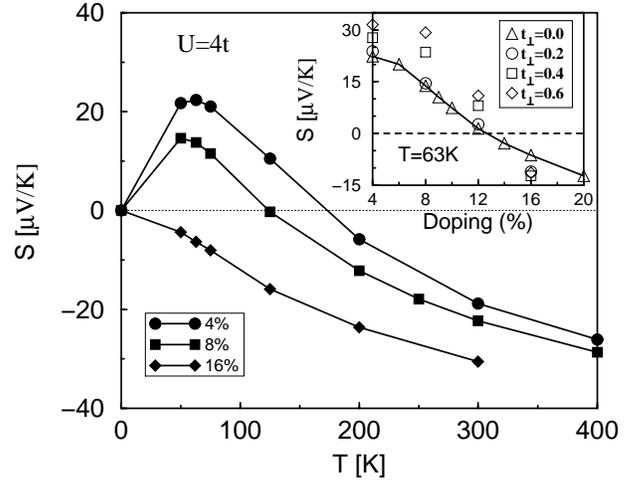, width=8.5cm}}
\caption{\label{fig:ybco-tep} 
TEP for the YBCO parameters as a
function of temperature for different hole dopings (inset:
doping dependence).
}
\end{figure}

The inset in Fig.~\ref{fig:ybco-tep} displays our results for the doping
dependence of the planar contribution to the TEP 
of YBCO which is also generic for other 
cuprates, using   $t'=-0.38 t$ and
$t''=-0.06 t$
 such that 
the Fermi surface is closed around the $(\pi, \pi)$ point
 (see Fig.~\ref{fig:ako}(b)).
In contrast to the LSCO system and in agreement with the
general trends in the experiments, 
the thermopower   changes its sign 
as   function of the doping concentration. 
In order to understand the origin of this behavior, we plot in 
Fig.~\ref{fig:ybco-tep}  the TEP as a function of temperature
for  different doping values.
As in LSCO,  a maximum of the TEP occurs   for small doping and small
temperatures.
However, for increasing $T$, the sign of $S$ becomes negative.
For larger doping, $S$ decreases until it becomes negative at all
temperatures.
In agreement with the experimental observation~\cite{BernhardTallon}, 
we find that the
negative slope of the TEP for large $T$ is almost doping independent. 
These phenomena can also be understood in terms of the pronounced 
momentum dependence of the scattering rates caused by antiferromagnetic
spin fluctuations.
We find that for low doping, the scattering rates are maximal 
for momentum states  near ${\vec k}_{\rm F} + {\vec Q}$.
However, as can be seen in Fig.~\ref{fig:ako}(b), 
these states are solely 
outside the Fermi surface
for ${\vec k}_{\rm F} + {\vec Q}$ near $(\frac{\pi}{2},\frac{\pi}{2})$, while they are
within the Fermi surface for ${\vec k}_{\rm F} + {\vec Q}$ near $(\pi,0)$,
in contrast to the LSCO system.
If these two regions in momentum space were of equal importance
for the suppression of near Fermi surface states in $\sigma(\omega)$,
they would   cancel each other, leading to a negative $S$, as found for
$U=0$. However,  the transport properties are dominated by the
 quasi-particles near the  diagonal of the 
Brillouin zone~\cite{HlubinaRice}. This can also be observed in the spectral function
plot for YBCO in Fig.~\ref{fig:ako}(d), in which the states near
$(\pi/2,\pi/2)$ are much more coherent than the ones near $(\pi,0)$.
For ${\vec k}_{\rm F}$ near $(\pi/2,\pi/2)$, 
  the  holes near   the Fermi surface are 
closer to ${\vec k}_{\rm F} + {\vec Q}$ than the  corresponding electron 
states and it follows that $S>0$   for sufficiently strong 
antiferromagnetic correlations. 
 Finally, due to the more subtle competition of different 
regions in momentum space for the YBCO parameters, 
the weakening of antiferromagnetic correlations
for increasing doping or temperature leads to the more rapid 
(as compared to LSCO) 
transition to  negative  values of the TEP, shown in Fig.~\ref{fig:ybco-tep}.
For a quantitative comparison of our results in
Fig.~\ref{fig:ybco-tep} with experiment,
it should be noted that the magnitude of $S$ as well as the 
doping level at which  $S$ changes its sign for low $T$ are smaller
than in  experiment.
One    reason  for this quantitative difference    might be the neglect
of the bilayer coupling in YBCO which was shown to enhance  the in-plane
antiferromagnetic correlations~\cite{HetzelHanke,YbcoBilayer}. 
By taking  bilayer coupling into account via an interlayer hopping $t_\perp$,
we solved  the FLEX equations and 
determined the   TEP. Note, that $t_\perp=0.4t$ is a reasonable
value for the interlayer hopping in YBCO~\cite{YbcoBilayer}.
The results are also shown in  the inset of Fig.~\ref{fig:ybco-tep}
for different
values of $t_\perp$ and demonstrate that indeed a considerable increase
of the absolute magnitude of $S$ occurs  due to 
inter-layer spin correlations.

Concerning the influence of the bandstructure
parametrization used for YBCO, it might be argued that
the velocity $v_{{\vec k},\alpha }$ on the Fermi surface
depends more strongly on the
direction of ${\vec k}$ than for other parametrizations 
(e.g. $t'=-0.45t$ and $t''=0$\cite{SP96}). 
However this is not essential for our results, since the
agreement with experiments would be even better if we were to  neglect this
${\vec k}$ dependence, which suppresses $S$.
We have used this parametrization since it yields better agreement
of the calculated spin susceptibility with 
the spin-excitation spectrum deduced from
NMR-experiments~\cite{BP}.

In summary, we have shown that the anomalous features of the in-plane
thermopower can be understood as consequences of short-range
antiferromagnetic correlations. The  opening of a pseudogap and the 
anomalous momentum dependence of the single-particle scattering rate
provide an  explanation of the general trends of  both
temperature and doping dependence of the in-plane thermopower.
This demonstrates that the thermoelectric power is a sensitive
probe of the anomalous nature of the low-energy excitations, caused by 
the antiferromagnetic correlations of the cuprates.

%
%
We are indebted  to E.~Arrigoni, K.~H.~Bennemann and B.~Stojkovi\'c 
for useful discussions.
Financial support by the Bavarian High-$T_c$ Program FORSUPRA is 
acknowledged.

\end{document}